\documentclass[aps,reprint,superscriptaddress,article]{revtex4-2}
\usepackage{graphicx,graphics,color,epsfig}
\usepackage{times}
\usepackage{epstopdf}
\usepackage{bm}
\usepackage{bbm}
\usepackage{amsmath}
\usepackage{amssymb}
\usepackage{diagbox}
\usepackage{color}
\usepackage{appendix}
\usepackage{dcolumn}
\usepackage[colorlinks=true, letterpaper=true, pdfstartview=FitV, linkcolor=blue, citecolor=blue, urlcolor=blue]{hyperref}

\begin{document}

\title{Magic of nonlocal geometric force: lighting up optical transition and transporting angular momentum by chiral phonons}

\author{Hao Chen}\thanks{These authors contributed equally to this work.}
\affiliation{Department of Physics, University of Science and Technology of China, Hefei 230026, China}
\affiliation{School of Science, Nanjing University of Posts and Telecommunications, Nanjing 210023, China}

\author{Haoshu Li}\thanks{These authors contributed equally to this work.}
\affiliation{Department of Physics, University of Science and Technology of China, Hefei 230026, China}

\author{Lifa Zhang}
\affiliation{School of Physics and Technology, Nanjing Normal University, Nanjing 210023, China}

\author{Qian Niu}
\affiliation{Department of Physics, University of Science and Technology of China, Hefei 230026, China}

\begin{abstract}

We investigate the impact of the nonlocal geometric force---arising from the molecular Berry curvature---on the lattice dynamics of magnetic materials with broken time-reversal symmetry. A first-principles computational framework is established to evaluate this force across the entire Brillouin zone. We apply it to monolayer CoCl$_2$, a ferromagnetic half-semiconductor with a narrow bandgap forbidding direct dipolar optical transition. At the phonon Brillouin zone center, the pronounced nonlocal geometric force leads to a splitting of the two upper optical phonon branches by $3 \times 10^{-2}$ THz, transforming the phonons into chiral modes. Optical chiral phonons can light up the intravalley dark exciton via absorpting circularly polarized photons. Furthermore, acoustic chiral phonons induced by the nonlocal geometric force can transport angular momentum and contribute to a non-dissipative phonon Hall viscosity.

\end{abstract}

\maketitle

\textcolor{blue}{\textit{Introduction.} ---}
When a magnetic field is applied to some specific materials, the splitting of the phonon spectrum at the $\Gamma$ point is observed. This phenomenon give rise to a large effective phonon magnetic moment~\cite{cheng2020large,ren2021phonon,zhang2023gate,wu2023fluctuation}. The reason of phonon splitting is that the application of magnetic field destroys the time reversal $\mathcal{T}$ symmetry of the system. However, beyond external fields, the intrinsic magnetic properties of materials can also break $\mathcal{T}$ symmetry. While the effects of material magnetism on electronic systems are well-studied, including phenomena like the anomalous Hall effect~\cite{jungwirth2002anomalous,nagaosa2010anomalous,chen2014anomalous,nakatsuji2015large,liang2018anomalous,mciver2020light} and the quantum anomalous Hall effect~\cite{laughlin1983anomalous,yu2010quantized,chang2013experimental,deng2020quantum,chang2023colloquium}, its impact on phonons remains largely unexplored, due to a lack of research methodologies addressing $\mathcal{T}$ symmetry breaking in phonon systems.

On the other hand, the Born-Oppenheimer approximation assumes adiabatic evolution of electron states with ion motion~\cite{born1996dynamical}. During this evolution, the electron ground state can accumulate non-trivial geometric phases in the absence of $\mathcal{T}$ symmetry~\cite{mead1979determination,mead1992geometric}, which is referred to as the molecular Berry phase~\cite{min2014molecular}. The associated nonlocal geometric force induced by molecular Berry curvature can influence lattice dynamics, suggesting that $\mathcal{T}$ symmetry breaking in electronic systems could propagate into phonon systems via it. Previous related studies focused on model study~\cite{saparov2022lattice} or first-principles study limited on the $\Gamma$ point~\cite{PhysRevLett.130.086701}, and so far a general study is still lacking.

In this letter, we present a first-principles approach to calculate the nonzero molecular Berry curvature in real magnetic materials, rather than in idealized models. This method enables the incorporation of $\mathcal{T}$ symmetry breaking in the lattice dynamics of magnetic materials and can be applied to the entire Brillouin zone. We apply it to the two-dimensional (2D) ferromagnetic material CoCl$_2$, a semiconductor with a narrow bandgap. Our calculations reveal a significant molecular Berry curvature near the $\Gamma$ point of the phonon Brillouin zone, primarily originating from the $M$ point of the electronic Brillouin zone. The nonlocal geometric force caused by molecular Berry curvature lifts the degeneracy of the seventh and eighth phonon branches at the $\Gamma$ point, resulting in a frequency splitting of $3 \times 10^{-2}$ THz, measurable by existing optical techniques. As the phonon branches split, the phonon modes acquire chirality and nonzero pseudoangular momentum (PAM). These chiral phonons can interact with electrons during intravalley transitions, enabling the optical excitation to light up dark excitons. Additionally, we demonstrate that the nonlocal geometric force gives rise to a finite Berry curvature and angular momentum in acoustic phonons, thereby generating a obvious phonon angular momentum Hall effect and a macroscopic Hall viscosity. Our work provides a first-principles method to investigate the effects of material magnetism on lattice dynamics and furthers the understanding of phonon chirality phenomena in magnetic materials.

\begin{figure}[htbp]
\includegraphics[width=3.2 in,angle=0]{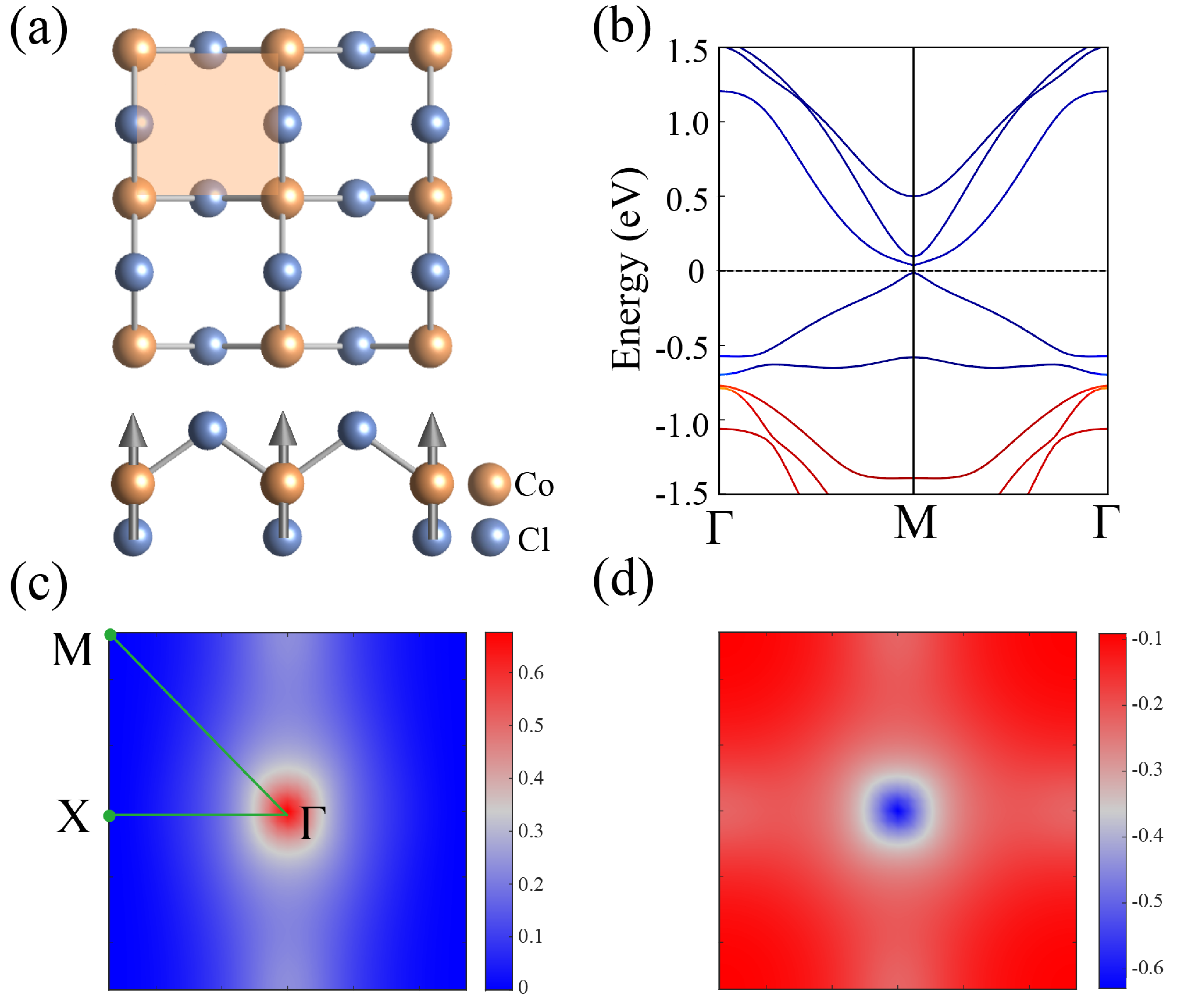}
\caption{\label{fig1} (a) Top and side views of monolayer CoCl$_2$ structure, the arrows indicate the magnetic moments of Co atoms. (b) Electron band structure of CoCl$_2$ with spin-orbit coupling included. The color indicates the spin polarization of the electron states, blue represents spin-down and red represents spin-up. (c)-(d) Calculated molecular Berry curvature for different components $G_{\text{Co} y}^{\text{Cl}_{2}x}$ and $G_{\text{Co} y}^{\text{Co} x}$, the unit is {\AA}$^{-2}$.}
\end{figure}

\textcolor{blue}{\textit{Nonlocal geometric force from first principles calculations.} ---}
We select CoCl$_2$ as the material of interest, with an optimized lattice constant of $4.0${\AA}. The unit cell contains two Cl atoms and one Co atom, with the Co atoms magnetic moment aligned consistently, indicating that the material is ferromagnetic, as shown in Fig.~\ref{fig1}(a). Viewed from above, the material forms a square lattice with space group No.115 and $C_4\mathcal{I}$ rotoinversion symmetry (it transforms pseudovectors in the same way as $C_4$ rotational symmetry, since spatial inversion symmetry $\mathcal{I}$ preserves pseudovectors). In the side view, the three atoms in the unit cell are not coplanar, resulting in a buckled monolayer structure that breaks $\mathcal{I}$ but retains mirror symmetries $M_x$ and $M_y$.

We obtained the electron band structure of CoCl$_2$ through first-principles calculations, with the spin components indicated on the bands. The spin-down states are shown in blue, while the spin-up states are represented in red, as shown in Fig.~\ref{fig1}(b). From the diagram, we clearly observe that both the valence band maximum (VBM) and conduction band minimum (CBM) are located at the $M$ point and are in the spin-down state. The bandgap is narrow, measuring 0.02 eV, indicating that CoCl$_2$ is a direct-gap ferromagnetic semiconductor. Building on this, analogous to the definition of half-metals, this semiconductor may be termed a \emph{half-semiconductor} since its conduction and valence band electrons exhibit fully spin-polarized states with identical spin orientation. We calculated the phonon spectrum without considering the nonlocal geometric force and found that the phonon branches are degenerate at high symmetry points $\Gamma$ and $M$. Moreover, the phonon modes vibrate linearly and do not have chirality. Relevant information is included in the Supplementary Material~\cite{supplemental}.

We further introduce the concept of molecular Berry curvature. In the Born-Oppenheimer approximation, the electrons are assumed to reside in their instantaneous ground state $|\Phi_{0}(\{\bm R\})\rangle$ for a given crystal structure $\{\bm R\}$. As the lattice structure evolves, the electron ground state adiabatically adjusts, accumulating geometric phases that influence the lattice dynamics. The Hamiltonian of the crystal lattice is expressed as~\cite{saparov2022lattice}
\begin{equation}\label{HL}
H_{L}=\sum\limits_{l,\kappa}\frac{1}{2M_{\kappa}}(\bm p_{l,\kappa}-\hbar \bm A_{l,\kappa}(\{\bm R\}))^{2}+V_{\text{eff}}(\{\bm R\}),
\end{equation}
where $\bm A_{l,\kappa}(\{\bm R\})=i\langle \Phi_{0}(\{\bm R\})|\nabla_{l,\kappa}\Phi_{0}(\{\bm R\}) \rangle$ is the molecular Berry connection that characterizes the geometric phase of the electronic ground state. Here, $\bm p_{l,\kappa}=-i\hbar \nabla_{l,\kappa}$ represents the canonical momentum of the $\kappa$th atom in the $l$th unit cell, and $V_{\text{eff}}(\{\bm R\})$ denotes the effective scalar potential. The molecular Berry connection can lead to a gauge invariant molecular Berry curvature:

\begin{equation}\label{G}
G_{\kappa^{'}\beta}^{\kappa\alpha}(\bm R_{l},\bm R_{l^{'}})=2\text{Im}  \left\langle \frac{\partial\Phi_{0}}{\partial R_{l',\kappa'\beta}} \Big| \frac{\partial\Phi_{0}}{\partial R_{l,\kappa\alpha}} \right\rangle,
\end{equation}
where $\alpha, \beta$ are the components of the Cartesian coordinates. For more calculation details, please refer to the Supplementary Material~\cite{supplemental}.

Figs.~\ref{fig1}(c) and (d) display the molecular Berry curvature for various components. Specifically, $G_{\text{Co} y}^{\text{Cl}_{2}x}$ represents the effect of the second Cl atom moving in the $x$-direction on the Co atom in other unit cells in the $y$-direction. Similarly, $G_{\text{Co} y}^{\text{Co} x}$ indicates the effect of the movement of Co atoms in one unit cell along the $x$-direction on the Co atoms in other unit cells in the $y$-direction. Other components are provided in the Supplementary Material~\cite{supplemental}. The largest molecular Berry curvature is concentrated around the $\Gamma$ point, with magnitudes reaching up to $10^{-1}$ {\AA}$^{-2}$. By performing a Fourier transform back to real space to evaluate $G_{\kappa^{'}\beta}^{\kappa\alpha}(\bm R_{l},\bm R_{l^{'}})$ in Eq.~(\ref{G}), since $G_{\kappa^{'}\beta}^{\kappa\alpha}(\bm{k})$ is concentrated near the center of the Brillouin zone, the corresponding real-space distribution exhibits a broad peak. This broad distribution implies that the motion of one atom can be influenced by the force of another atom far away over distances on the order of 10-100{\AA}. This effect manifests as a nonlocal geometric force, which contrasts with the conventional Raman spin-lattice coupling model. In that model, the spin-lattice coupling term is momentum independent, implying uniformity across the entire Brillouin zone and corresponding to a local force in real space. Furthermore, the conventional model predicts a splitting of the acoustic branches at the $\Gamma$ point, which contradicts experimental observations and relies on an external magnetic field. In contrast, our approach based on nonlocal geometric force reproduces the correct behavior without requiring any external field.

The distribution of molecular Berry curvature is closely related to the electron band structure. Since both the VBM and CBM are located at the $M$ point, the simplest phonon-induced virtual electron transition does not require the phonon to provide momentum, corresponding to the $\Gamma$ point of the phonon Brillouin zone.

\begin{figure}[htbp]
\includegraphics[width=1\linewidth]{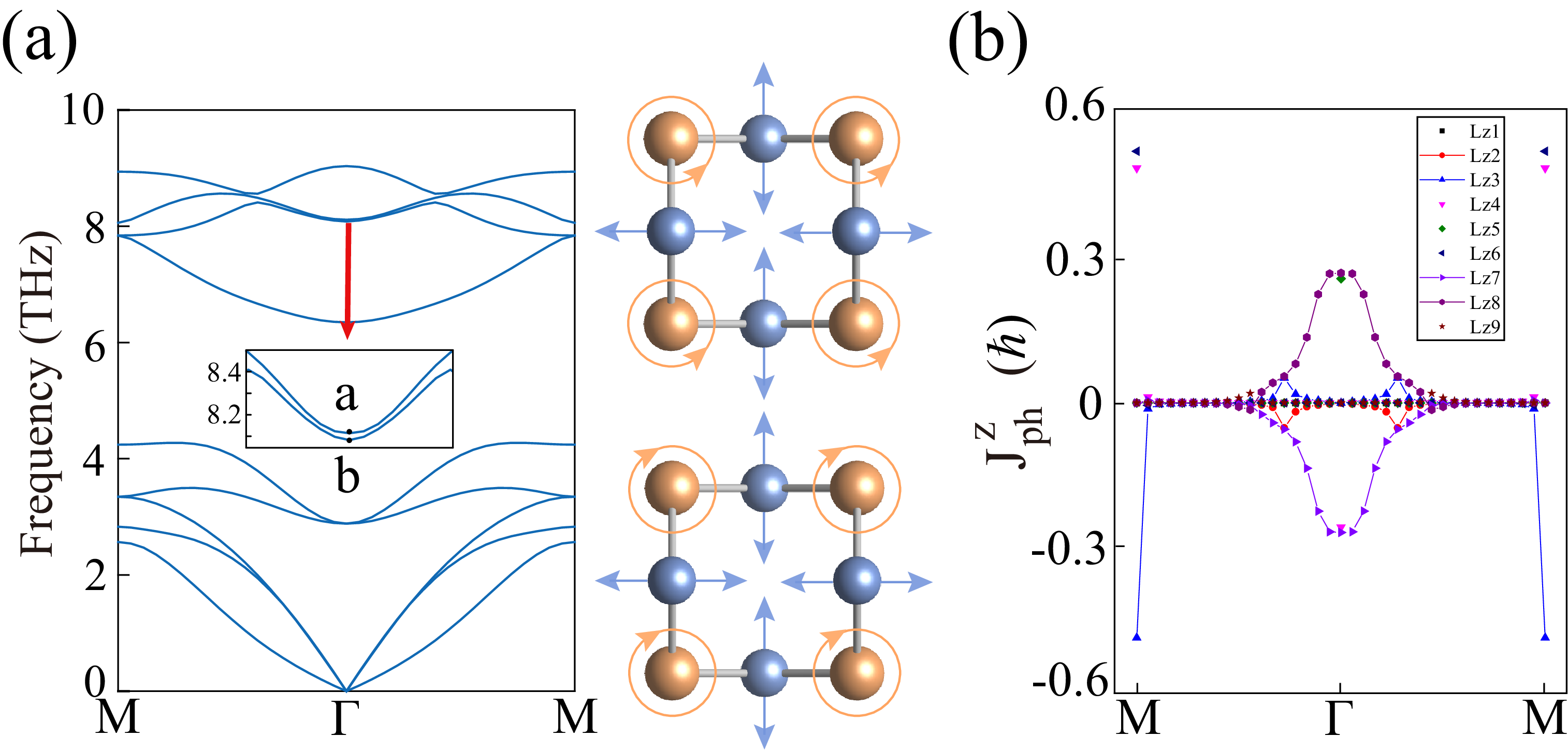}
\caption{\label{fig2}(a) Phonon spectrum with molecular Berry curvature. The molecular Berry curvature lifts the degeneracy of phonon branches at the high-symmetry points. We focus on the larger splitting between the seventh and eighth branches at the $\Gamma$ point; the inset presents a magnified view of the phonon splitting and the corresponding vibrational modes $a$ and $b$, with the vibration trajectories shown on the right. (b) The angular momentum $J_{\text{ph}}^{z}$ for each branch along the high-symmetry path.}
\end{figure}

\textcolor{blue}{\textit{Chiral phonon splitting from nonlocal geometric force.} ---}
Consistent with the distribution of the molecular Berry curvature in the phonon Brillouin zone in Figs.~\ref{fig1} (c) and (d), the larger phonon spectrum intrinsic splitting occurs at the $\Gamma$ point, especially the splitting of the seventh and eighth branches is the most obvious, as shown in Fig.~\ref{fig2} (a). Our calculations indicate that this splitting can reach $3\times10^{-2}$ THz, which corresponds to a huge effective magnetic field. As it splits, the higher-frequency phonon mode $a$ has a right-handed character, while the lower-frequency phonon mode $b$ has a left-handed character. To analyze these chiral vibrational modes clearly, we plotted the vibrational trajectories on the right side of Fig.~\ref{fig2} (a). The image shows that the Co atom in phonon mode $a$ moves counterclockwise, while the Co atom in phonon mode $b$ performs clockwise chiral vibration. The Cl atoms in both phonon modes vibrate linearly. From Fig.~\ref{fig1}, we also notice that although the molecular Berry curvature values outside the $\Gamma$ point are smaller than those at the $\Gamma$ point, they are not zero. Our calculation results show that the seventh and eighth branches have a splitting of $8\times10^{-4}$ THz at the $M$ point, which is also accompanied by the appearance of chiral phonons.

Due to the chiral vibration of atoms, phonons have angular momentum, defined as $\bm J_{\text{ph}}=\sum\limits_{l\kappa}\bm u_{l\kappa}\times \dot{\bm u}_{l\kappa}$~\cite{zhang2014angular}. For a 2D system, the $z$ component of the angular momentum becomes $\bm J_{\text{ph}}^{z}=\sum\limits_{l\kappa}(\bm u_{l\kappa}^{x} \bm \dot{u}_{l\kappa}^{y} - \bm u_{l\kappa}^{y} \bm \dot{u}_{l\kappa}^{x})$.   Fig.~\ref{fig2} (b) shows the $z$-direction phonon angular momentum of the each branches along the $M-G-M$ path. This shows that our method is not limited to the $\Gamma$ point, but the influence of nonlocal geometric force on phonon dynamics in the entire Brillouin zone can be calculated.

Next we focus on the chiral phonons at the $\Gamma$ point. Since the system has $C_4$ rotational symmetry, the chiral phonons at Brillouin zone center can define the quantized PAM, the definition is~\cite{zhang2015chiral,wang2022chiral}
\begin{equation}\label{PAM}
\mathcal{R}[(\pi/2),z]|u_{\bm k,\nu}\rangle= e^{-i(\pi/2)\ell_{\text{ph}}^{\bm k,\nu}} |u_{\bm k,\nu}\rangle,
\end{equation}
where $\mathcal{R}[(\pi/2),z]$ is the fourfold rotational operator along the $z$ direction, $u_{\bm k,\nu}$ is the phonon wavefunction and $\ell_{\text{ph}}^{\bm k,\nu}$ is the phonon PAM. According to this formula, we obtain the PAM of the two phonon modes $a$ and $b$ as $+1$ and $-1$ respectively. This makes these chiral phonons obey the selection rule in their coupling process with photons and electrons. For the Raman scattering process, the conservation of energy, momentum and PAM need to be satisfied. The Raman shift is equal to the energy of the observed phonon mode. We mentioned above that the frequency gap of the $a$ and $b$ modes is $3\times 10^{-2}$ THz, which makes it easy for Raman detection to distinguish phonon energy between the two modes. For PAM, this scattering process satisfies the relation $\ell_{i}- \ell_{s} =\ell_{\text{ph}}$, where $\ell_{i}$($\ell_{s}$) represents the circularly polarized of the incident (scattered) light. For a Raman process where the incident light is right-handed ($\ell_{i}=+1$) and the scattered light is left-handed ($\ell_{s}=-1$), only left-handed phonons ($\ell_{\text{ph}}=-1$) can satisfy the PAM conservation of this process. On the contrary, for the Raman process where the incident light is left-handed and the scattered light is right-handed, only right-handed phonons can satisfy the conservation of PAM. Therefore, when the incident/scattered light polarization is $\sigma^{+}/\sigma^{-}$ ($\sigma^{-}/\sigma^{+}$), the $b$ ($a$) phonon mode can be selectively detected.

\begin{figure}[htbp]
\includegraphics[width=2.8 in,angle=0]{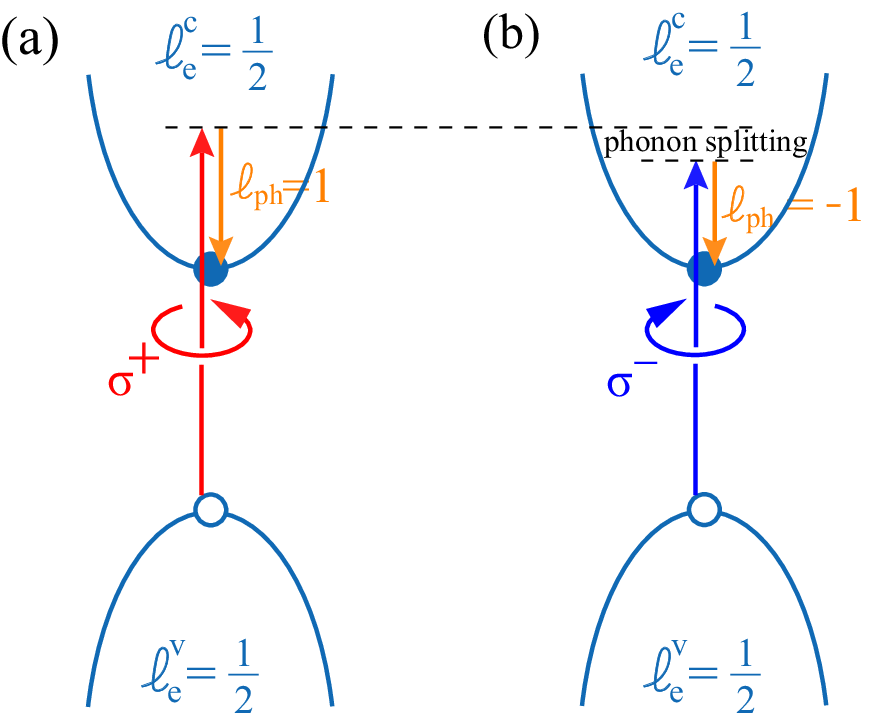}
\caption{\label{fig3} Schematic diagram of electron intravalley transition. (a) An electron can absorb a right circularly polarized photon $\sigma^{+}$ by emitting a right-handed phonon $\ell_{\text{ph}}=+1$. (b) A left circularly polarized photon $\sigma^{-}$ is absorbed by emitting a left-handed phonon $\ell_{\text{ph}}=-1$. $\ell_e^{v}$ is the electron PAM for VBM and $\ell_e^{c}$ is the electron PAM for CBM.}
\end{figure}

\textcolor{blue}{\textit{Optical chiral phonons light up intravalley dark excitons.} ---}
In most low-dimensional direct bandgap semiconductor systems, electron-hole excitations (excitons) near the band edge can be divided into two categories, depending on whether their total angular momentum is $\ell_z=\pm1$ or $\ell_z=\pm2, 0$. According to the standard selection rules for optical transitions, the excitons with $\ell_z=\pm 1$ can be excited by photons or recombine by emitting photons, thus being referred to as bright excitons. Instead, excitons with $\ell_z=\pm2, 0$ are called dark excitons because they cannot effectively couple to the radiation field. We calculated the electron PAM at the VBM and CBM, both of which are equal to $\frac{1}{2}$. The transition process has to obey the PAM conservation rule $\ell_{e}^C-\ell_{e}^V=\ell_\text{photon}$ where $\ell_{e}$ represents the PAM for the valley electronic states and $\ell_\text{photon}=\pm1$ represents right/left circularly polarized incident light ($\sigma^{+}/\sigma^{-}$). Since the electron PAM of both VBM and CBM bands is the same, the left side of the equation is zero implying the exciton state is dark. However, when chiral phonons participate in this transition process, the PAM conservation rule becomes
\begin{equation}
	\ell_{e}^C-\ell_{e}^V=\ell_\text{photon}-\ell_\text{ph}\quad \mod 4,
\end{equation}
the nonzero phonon PAM enables the electron transition to absorb photons with $\ell_\text{photon}=\pm1$, thereby lighting up the dark excitons in the valley. For the process emission right-handed phonons ($\ell_{\text{ph}}=+1$), the transition absorbs right circularly polarized light. For the process emission left-handed phonons ($\ell_{\text{ph}}=-1$), the transition absorbs left circularly polarized light.

As shown in Fig.~\ref{fig3}, energy conservation expressed as $\hbar(\omega_{\text{photon}}-\omega_{\text{ph}})=E_{gap}$, $E_{gap}$ is the energy of the electron band gap. This conservation implies that the energy difference between the right-handed phonon (mode $a$) and left-handed phonon (mode $b$) necessitates the use of incident light with different frequencies for the two processes. This difference allows one to distinguish between the two chiral phonon modes through the following procedure. Consider a linearly polarized incident light beam tuned to the frequencies corresponding to the processes depicted in Figs.~\ref{fig3}(a) and \ref{fig3}(b), respectively. For the frequency associated with Fig.~\ref{fig3}(a), right circularly polarized photons are preferentially absorbed, resulting in predominantly left polarized transmitted light. Conversely, for the frequency corresponding to Fig.~\ref{fig3}(b), left circularly polarized photons are preferentially absorbed, and the transmitted light exhibits right polarization. Thus, the distinct absorption behaviors of linearly polarized light at different frequencies provide a means to selectively probe transitions involving either the right-handed or the left-handed phonon. This can manifest as optical circular dichroism, which can be experimentally detected. Compared to previous works~\cite{zhu2018observation}, this detection proposal is based on a single process mechanism, achieving operational simplicity by strategically employing intravalley electronic transitions previously unexplored mechanism.

\textcolor{blue}{\textit{Transport phenomenon dominated by acoustic chiral phonons} ---}
Having discussed the impact of the nonlocal geometric force on the properties of optical phonons, we now turn to its influence on acoustic phonons. In contrast to our focus on microscopic optical excitation processes in the discussion of optical phonons, our attention here shifts to transport processes and macroscopic elastic properties.

First, we address transport processes. It is known that, due to the characteristics of the Bose distribution, thermal transport is predominantly governed by the properties of acoustic phonons. Under the influence of the nonlocal geometric force, acoustic chiral phonons acquire a nonzero anomalous velocity arising from the nonzero phonon Berry curvature. Consequently, this leads to the phonon angular momentum Hall effect. Under a temperature gradient $\nabla_\nu T$, the expected value of the phonon angular momentum current is given by $\langle j_\mu\rangle=-\beta_{\mu\nu} \nabla_\nu T$, where the Hall response coefficient for phonon angular momentum is defined as
\begin{align}
	\beta_{xy} = \sum_{i} \int \frac{d^2 k}{(2\pi)^2} \Omega^z_{\bm{k}i} (J^z_{\bm{k}i}/\hbar) s(\omega_{\bm{k},i}).
	\label{beta}
\end{align}
Here, $\Omega^z_{\bm{k}i}$ is the phonon Berry curvature, $J^z_{\bm{k}i}$ is the phonon angular momentum, and $s(\omega_{\bm{k},i}) = k_B [1+n_B(\hbar \omega_{\bm{k},i})]\text{log}[1+n_B(\hbar \omega_{\bm{k},i})]-n_B(\hbar \omega_{\bm{k},i})\text{log}[n_B(\hbar \omega_{\bm{k},i})]$ is the phonon entropy.

\begin{figure}[htbp]
\includegraphics[width=1\linewidth]{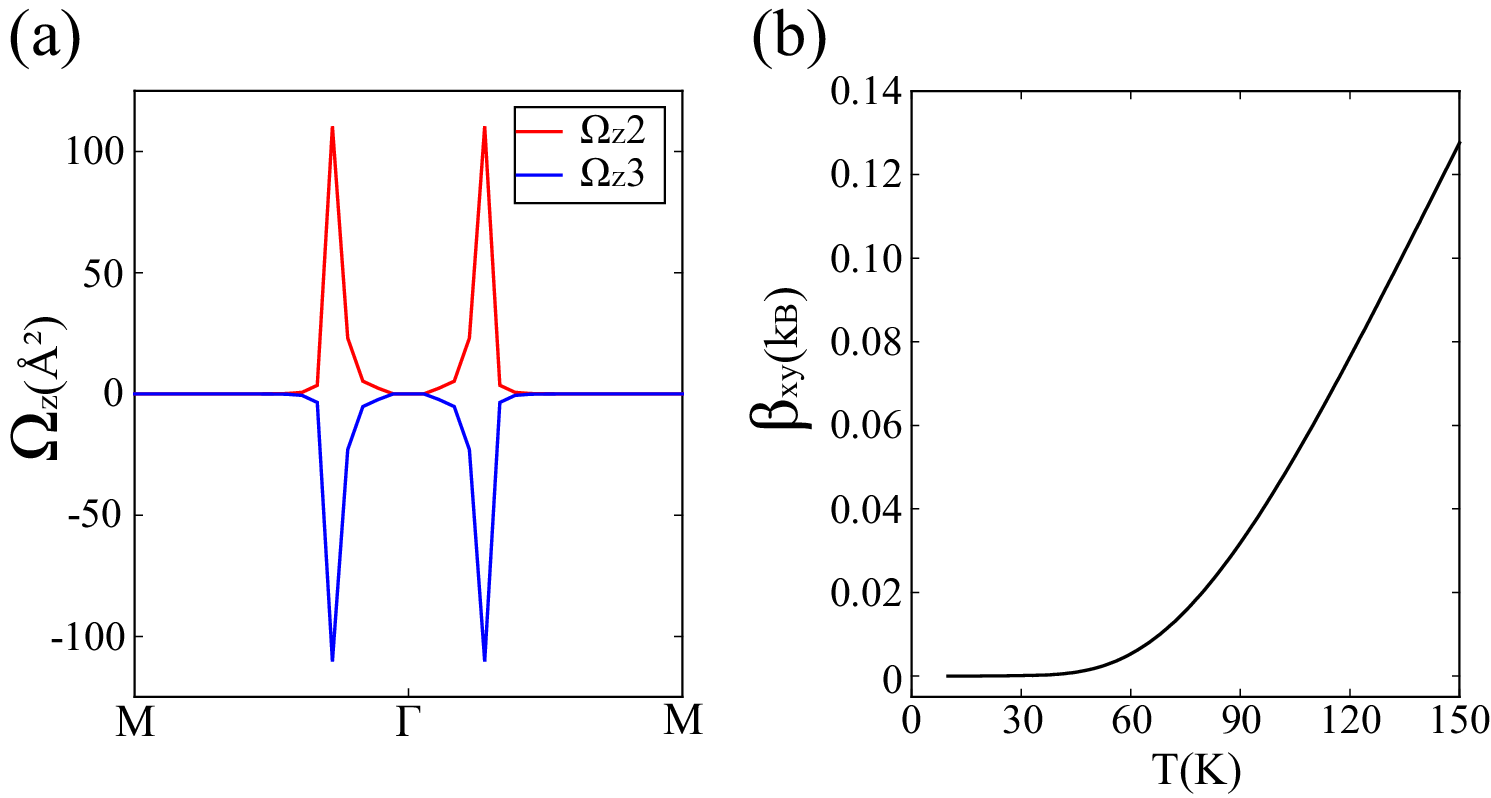}
\caption{\label{fig4} (a) Phonon Berry curvature of longitudinal and transverse acoustic branches. (b) Temperature dependence of the phonon angular momentum Hall conductivity. The unit is the Boltzmann constant $k_B$, where 1 $k_B$ indicates that, under a longitudinal temperature gradient of 1 $\text{K/m}$ applied to a square sample, angular momentum of $1.3 \times 10^{11} \hbar$ is transported in the transverse direction.}
\end{figure}

The near degeneracy of the longitudinal and transverse acoustic branches in this material gives rise to pronounced phonon Berry curvature (Fig.\ref{fig4}(a)) and significant angular momentum (Fig.\ref{fig2}(b)) across a wide Brillouin zone region. Their combined effect results in a large phonon angular momentum Hall response coefficient $\beta_{xy}$, as illustrated in Fig.~\ref{fig4}(b). The calculated phonon Berry curvature $\Omega_{\bm{k}i}^{z}$ and phonon angular momentum $J^z_{\bm{k}i}$ exhibit a quadratic dependence on momentum ($\propto k^2$). Through Eq.~(\ref{beta}), this scaling dictates $\beta_{xy}\propto T^6$ in the low-temperature limit. In the high-temperature limit, $\beta_{xy} = c \log\left(\frac{k_B T}{\epsilon_0}\right)$, with $c = -\sum_{i} \int \frac{d^2 k}{(2\pi)^2} \Omega^z_{\bm{k}i} (J^z_{\bm{k}i}/\hbar) \log\left(\frac{\hbar \omega_{\bm{k},i}}{\epsilon_0}\right)$ and $\epsilon_0$ denotes the maximum energy of all phonons.

Furthermore, we emphasize that one cannot obtain nonzero phonon Berry curvature or angular momentum when neglecting the effect of nonlocal geometric force. Under this assumption, the Hall response coefficient for phonon angular momentum would vanish. Therefore, the observed angular momentum Hall effect in the material directly reflects the influence of the nonlocal geometric force, providing experimental evidence for its existence. Since phonon angular momentum can drive magnetization, the transport of angular momentum can further induce magnetization changes at remote locations, offering promising applications in spintronics and magnetism research.

Apart from its role in chiral phonon transport, the nonlocal geometric force also influences low-energy elastic properties of acoustic waves. This force induces a dissipationless component in the viscosity tensor within the low-energy regime of acoustic lattice dynamics - a phenomenon termed phonon Hall viscosity \cite{PhysRevLett.131.236301}. The emergent viscosity governs the viscoelastic stress response to external strain rates through the constitutive relation $\sigma_{\lambda \mu} = \eta_{\lambda \mu \nu \xi} \dot{\epsilon}_{\nu \xi}$, where $\sigma_{\lambda \mu}$ denotes the stress tensor, $\epsilon_{\nu \xi}$ the strain tensor, and the Hall viscosity $\eta^H \equiv \eta_{xxxy}$ corresponds to a part of the viscosity tensor $\eta_{\lambda \mu \nu \xi}$. Within the basis of the LA and TA modes and along the high-symmetry line, the modified dynamic matrix takes the form:
\begin{align}
	D(\bm{k},\omega) = \begin{pmatrix}
		v_L^2 \bm{k}^2 & i \eta^H \omega \bm{k}^2 \\
		-i \eta^H \omega \bm{k}^2 & v_T^2 \bm{k}^2
	\end{pmatrix},
	\label{eq: dy_vis}
\end{align}
where $v_L$ and $v_T$ represent the longitudinal and transverse acoustic velocities, respectively. Meanwhile, the microscopic lattice dynamics modified by the nonlocal geometric force is $\omega^2 u_{\bm{k}}=D_{\bm{k}} u_{\bm{k}} + 2 i \omega \tilde{G}_{\bm{k}} u_{\bm{k}}$. This microscopic equation becomes Eq.~(\ref{eq: dy_vis}) at the low-energy and long-wavelength limit, and it implies two distinct mechanisms contributing to the Hall viscosity $\eta^H$:
Intra-acoustic contribution $\eta^H_{AA}$ which is from direct molecular Berry curvature couplings between longitudinal and transverse acoustic modes; Acoustic-optical contribution $\eta^H_{AO}$ which is an indirect contribution arising from integrating out high-energy optical modes through acoustic-optical molecular Berry curvature couplings. For the material under investigation, the respective contributions are quantified as $\eta^H_{AA}\sim 1.43\times10^{-24}\text{kg}\,\text{s}^{-1}$ and $\eta^H_{AO}\sim-5.74\times10^{-39}\text{kg}\,\text{s}^{-1}$. Based on this computational result, it can be concluded that in contrast to previous studies examining extrinsic magnetism \cite{PhysRevLett.131.236301}, the Hall viscosity arising from intrinsic magnetism primarily originates from intra-acoustic contributions rather than acoustic-optical contributions.

\textcolor{blue}{\textit{Discussion and conclusion.} ---}
The magnetic moment of ferromagnetic materials can be flipped by an external magnetic field, so the spin polarization of the electron band will also be flipped. This indicate that the sign of the molecular Berry curvature will be reversed. Analogous to the directional reversal of the nonlocal effective magnetic field, wherein the angular momentum of the phonon mode is inverted. Using this ferromagnetic mechanism, we can control chiral phonons through external magnetic field. Even if the magnetic field is removed, the chirality of the phonons can be well preserved. The directions of phonon angular momentum Hall current can also be artificially controlled. Moreover, since these acoustic phonon modes with opposite chirality have a group velocity difference, the phonon Faraday effect can be generated.

In summary, the method we developed allows us to obtain the molecular Berry curvature in magnetic materials using first-principles calculations, which can induce the $\mathcal{T}$ symmetry breaking of electronic states into the phonon dynamics. We applied the method to a 2D ferromagnetic material CoCl$_2$, where the direct narrow bandgap at $M$ point makes the molecular Berry curvature at the $\Gamma$ point of the phonon Brillouin zone relatively large. This results in the seventh and eighth phonon branches of the $\Gamma$ point being opened up to a degeneracy of $3\times 10^{-2}$ THz, accompanied by the appearance of chiral optical phonons. The chirality and energy splitting of these phonons can be well detected by helicity-resolved Raman scattering. In addition, these chiral phonons at the $\Gamma$ point can also participate in the intravalley transition of electrons, and the nonzero phonon PAM can enable the transition absorb circularly polarized photons, thus lighting up the intravalley dark exciton. Our analysis also reveals that the nonlocal geometric force fundamentally alters acoustic phonon dynamics and induces finite phonon Berry curvature and angular momentum. This gives rise to an angular momentum Hall effect, where the Hall response is markedly enhanced due to the nearly degenerate longitudinal and transverse acoustic branches. Moreover, this nonlocal geometric force also generates a finite Hall viscosity, primarily arising from the coupling between acoustic branches.

\textcolor{blue}{\textit{Acknowledgement} ---}
This work was supported by the National Natural Science Foundation of China (12234017), National Key R\&D Program of China (2023YFA1407001), and Department of Science and Technology of Jiangsu Province (BK20220032). Hao Chen acknowledges support from the Natural Science Research Start-up Foundation of Recruiting Talents of Nanjing University of Posts and Telecommunications (Grant No.NY224105). Haoshu Li acknowledges support from the China Postdoctoral Science Foundation 2023M743398.

\bibliography{aps-main}

\begin{thebibliography}{27}%
\makeatletter
\providecommand \@ifxundefined [1]{%
 \@ifx{#1\undefined}
}%
\providecommand \@ifnum [1]{%
 \ifnum #1\expandafter \@firstoftwo
 \else \expandafter \@secondoftwo
 \fi
}%
\providecommand \@ifx [1]{%
 \ifx #1\expandafter \@firstoftwo
 \else \expandafter \@secondoftwo
 \fi
}%
\providecommand \natexlab [1]{#1}%
\providecommand \enquote  [1]{``#1''}%
\providecommand \bibnamefont  [1]{#1}%
\providecommand \bibfnamefont [1]{#1}%
\providecommand \citenamefont [1]{#1}%
\providecommand \href@noop [0]{\@secondoftwo}%
\providecommand \href [0]{\begingroup \@sanitize@url \@href}%
\providecommand \@href[1]{\@@startlink{#1}\@@href}%
\providecommand \@@href[1]{\endgroup#1\@@endlink}%
\providecommand \@sanitize@url [0]{\catcode `\\12\catcode `\$12\catcode
  `\&12\catcode `\#12\catcode `\^12\catcode `\_12\catcode `\%12\relax}%
\providecommand \@@startlink[1]{}%
\providecommand \@@endlink[0]{}%
\providecommand \url  [0]{\begingroup\@sanitize@url \@url }%
\providecommand \@url [1]{\endgroup\@href {#1}{\urlprefix }}%
\providecommand \urlprefix  [0]{URL }%
\providecommand \Eprint [0]{\href }%
\providecommand \doibase [0]{https://doi.org/}%
\providecommand \selectlanguage [0]{\@gobble}%
\providecommand \bibinfo  [0]{\@secondoftwo}%
\providecommand \bibfield  [0]{\@secondoftwo}%
\providecommand \translation [1]{[#1]}%
\providecommand \BibitemOpen [0]{}%
\providecommand \bibitemStop [0]{}%
\providecommand \bibitemNoStop [0]{.\EOS\space}%
\providecommand \EOS [0]{\spacefactor3000\relax}%
\providecommand \BibitemShut  [1]{\csname bibitem#1\endcsname}%
\let\auto@bib@innerbib\@empty
\bibitem [{\citenamefont {Cheng}\ \emph {et~al.}(2020)\citenamefont {Cheng},
  \citenamefont {Schumann}, \citenamefont {Wang}, \citenamefont {Zhang},
  \citenamefont {Barbalas}, \citenamefont {Stemmer},\ and\ \citenamefont
  {Armitage}}]{cheng2020large}%
  \BibitemOpen
  \bibfield  {author} {\bibinfo {author} {\bibfnamefont {B.}~\bibnamefont
  {Cheng}}, \bibinfo {author} {\bibfnamefont {T.}~\bibnamefont {Schumann}},
  \bibinfo {author} {\bibfnamefont {Y.}~\bibnamefont {Wang}}, \bibinfo {author}
  {\bibfnamefont {X.}~\bibnamefont {Zhang}}, \bibinfo {author} {\bibfnamefont
  {D.}~\bibnamefont {Barbalas}}, \bibinfo {author} {\bibfnamefont
  {S.}~\bibnamefont {Stemmer}},\ and\ \bibinfo {author} {\bibfnamefont
  {N.}~\bibnamefont {Armitage}},\ }\bibfield  {title} {\bibinfo {title} {A
  large effective phonon magnetic moment in a {Dirac} semimetal},\ }\href@noop
  {} {\bibfield  {journal} {\bibinfo  {journal} {Nano Lett.}\ }\textbf
  {\bibinfo {volume} {20}},\ \bibinfo {pages} {5991} (\bibinfo {year}
  {2020})}\BibitemShut {NoStop}%
\bibitem [{\citenamefont {Ren}\ \emph {et~al.}(2021)\citenamefont {Ren},
  \citenamefont {Xiao}, \citenamefont {Saparov},\ and\ \citenamefont
  {Niu}}]{ren2021phonon}%
  \BibitemOpen
  \bibfield  {author} {\bibinfo {author} {\bibfnamefont {Y.}~\bibnamefont
  {Ren}}, \bibinfo {author} {\bibfnamefont {C.}~\bibnamefont {Xiao}}, \bibinfo
  {author} {\bibfnamefont {D.}~\bibnamefont {Saparov}},\ and\ \bibinfo {author}
  {\bibfnamefont {Q.}~\bibnamefont {Niu}},\ }\bibfield  {title} {\bibinfo
  {title} {Phonon magnetic moment from electronic topological magnetization},\
  }\href@noop {} {\bibfield  {journal} {\bibinfo  {journal} {Phys. Rev. Lett.}\
  }\textbf {\bibinfo {volume} {127}},\ \bibinfo {pages} {186403} (\bibinfo
  {year} {2021})}\BibitemShut {NoStop}%
\bibitem [{\citenamefont {Zhang}\ \emph {et~al.}(2023)\citenamefont {Zhang},
  \citenamefont {Ren}, \citenamefont {Wang}, \citenamefont {Cao},\ and\
  \citenamefont {Xiao}}]{zhang2023gate}%
  \BibitemOpen
  \bibfield  {author} {\bibinfo {author} {\bibfnamefont {X.-W.}\ \bibnamefont
  {Zhang}}, \bibinfo {author} {\bibfnamefont {Y.}~\bibnamefont {Ren}}, \bibinfo
  {author} {\bibfnamefont {C.}~\bibnamefont {Wang}}, \bibinfo {author}
  {\bibfnamefont {T.}~\bibnamefont {Cao}},\ and\ \bibinfo {author}
  {\bibfnamefont {D.}~\bibnamefont {Xiao}},\ }\bibfield  {title} {\bibinfo
  {title} {Gate-tunable phonon magnetic moment in bilayer graphene},\
  }\href@noop {} {\bibfield  {journal} {\bibinfo  {journal} {Phys. Rev. Lett.}\
  }\textbf {\bibinfo {volume} {130}},\ \bibinfo {pages} {226302} (\bibinfo
  {year} {2023})}\BibitemShut {NoStop}%
\bibitem [{\citenamefont {Wu}\ \emph {et~al.}(2023)\citenamefont {Wu},
  \citenamefont {Bao}, \citenamefont {Zhou}, \citenamefont {Wang},
  \citenamefont {Sun}, \citenamefont {Wen}, \citenamefont {Wan},\ and\
  \citenamefont {Zhang}}]{wu2023fluctuation}%
  \BibitemOpen
  \bibfield  {author} {\bibinfo {author} {\bibfnamefont {F.}~\bibnamefont
  {Wu}}, \bibinfo {author} {\bibfnamefont {S.}~\bibnamefont {Bao}}, \bibinfo
  {author} {\bibfnamefont {J.}~\bibnamefont {Zhou}}, \bibinfo {author}
  {\bibfnamefont {Y.}~\bibnamefont {Wang}}, \bibinfo {author} {\bibfnamefont
  {J.}~\bibnamefont {Sun}}, \bibinfo {author} {\bibfnamefont {J.}~\bibnamefont
  {Wen}}, \bibinfo {author} {\bibfnamefont {Y.}~\bibnamefont {Wan}},\ and\
  \bibinfo {author} {\bibfnamefont {Q.}~\bibnamefont {Zhang}},\ }\bibfield
  {title} {\bibinfo {title} {Fluctuation-enhanced phonon magnetic moments in a
  polar antiferromagnet},\ }\href@noop {} {\bibfield  {journal} {\bibinfo
  {journal} {Nat. Phys.}\ }\textbf {\bibinfo {volume} {19}},\ \bibinfo {pages}
  {1868} (\bibinfo {year} {2023})}\BibitemShut {NoStop}%
\bibitem [{\citenamefont {Jungwirth}\ \emph {et~al.}(2002)\citenamefont
  {Jungwirth}, \citenamefont {Niu},\ and\ \citenamefont
  {MacDonald}}]{jungwirth2002anomalous}%
  \BibitemOpen
  \bibfield  {author} {\bibinfo {author} {\bibfnamefont {T.}~\bibnamefont
  {Jungwirth}}, \bibinfo {author} {\bibfnamefont {Q.}~\bibnamefont {Niu}},\
  and\ \bibinfo {author} {\bibfnamefont {A.}~\bibnamefont {MacDonald}},\
  }\bibfield  {title} {\bibinfo {title} {Anomalous {Hall} effect in
  ferromagnetic semiconductors},\ }\href@noop {} {\bibfield  {journal}
  {\bibinfo  {journal} {Phys. Rev. Lett.}\ }\textbf {\bibinfo {volume} {88}},\
  \bibinfo {pages} {207208} (\bibinfo {year} {2002})}\BibitemShut {NoStop}%
\bibitem [{\citenamefont {Nagaosa}\ \emph {et~al.}(2010)\citenamefont
  {Nagaosa}, \citenamefont {Sinova}, \citenamefont {Onoda}, \citenamefont
  {MacDonald},\ and\ \citenamefont {Ong}}]{nagaosa2010anomalous}%
  \BibitemOpen
  \bibfield  {author} {\bibinfo {author} {\bibfnamefont {N.}~\bibnamefont
  {Nagaosa}}, \bibinfo {author} {\bibfnamefont {J.}~\bibnamefont {Sinova}},
  \bibinfo {author} {\bibfnamefont {S.}~\bibnamefont {Onoda}}, \bibinfo
  {author} {\bibfnamefont {A.~H.}\ \bibnamefont {MacDonald}},\ and\ \bibinfo
  {author} {\bibfnamefont {N.~P.}\ \bibnamefont {Ong}},\ }\bibfield  {title}
  {\bibinfo {title} {Anomalous {Hall} effect},\ }\href@noop {} {\bibfield
  {journal} {\bibinfo  {journal} {Rev. Mod. Phys.}\ }\textbf {\bibinfo {volume}
  {82}},\ \bibinfo {pages} {1539} (\bibinfo {year} {2010})}\BibitemShut
  {NoStop}%
\bibitem [{\citenamefont {Chen}\ \emph {et~al.}(2014)\citenamefont {Chen},
  \citenamefont {Niu},\ and\ \citenamefont {MacDonald}}]{chen2014anomalous}%
  \BibitemOpen
  \bibfield  {author} {\bibinfo {author} {\bibfnamefont {H.}~\bibnamefont
  {Chen}}, \bibinfo {author} {\bibfnamefont {Q.}~\bibnamefont {Niu}},\ and\
  \bibinfo {author} {\bibfnamefont {A.~H.}\ \bibnamefont {MacDonald}},\
  }\bibfield  {title} {\bibinfo {title} {Anomalous {Hall} effect arising from
  noncollinear antiferromagnetism},\ }\href@noop {} {\bibfield  {journal}
  {\bibinfo  {journal} {Phys. Rev. Lett.}\ }\textbf {\bibinfo {volume} {112}},\
  \bibinfo {pages} {017205} (\bibinfo {year} {2014})}\BibitemShut {NoStop}%
\bibitem [{\citenamefont {Nakatsuji}\ \emph {et~al.}(2015)\citenamefont
  {Nakatsuji}, \citenamefont {Kiyohara},\ and\ \citenamefont
  {Higo}}]{nakatsuji2015large}%
  \BibitemOpen
  \bibfield  {author} {\bibinfo {author} {\bibfnamefont {S.}~\bibnamefont
  {Nakatsuji}}, \bibinfo {author} {\bibfnamefont {N.}~\bibnamefont
  {Kiyohara}},\ and\ \bibinfo {author} {\bibfnamefont {T.}~\bibnamefont
  {Higo}},\ }\bibfield  {title} {\bibinfo {title} {Large anomalous {Hall}
  effect in a non-collinear antiferromagnet at room temperature},\ }\href@noop
  {} {\bibfield  {journal} {\bibinfo  {journal} {Nature}\ }\textbf {\bibinfo
  {volume} {527}},\ \bibinfo {pages} {212} (\bibinfo {year}
  {2015})}\BibitemShut {NoStop}%
\bibitem [{\citenamefont {Liang}\ \emph {et~al.}(2018)\citenamefont {Liang},
  \citenamefont {Lin}, \citenamefont {Gibson}, \citenamefont {Kushwaha},
  \citenamefont {Liu}, \citenamefont {Wang}, \citenamefont {Xiong},
  \citenamefont {Sobota}, \citenamefont {Hashimoto}, \citenamefont {Kirchmann}
  \emph {et~al.}}]{liang2018anomalous}%
  \BibitemOpen
  \bibfield  {author} {\bibinfo {author} {\bibfnamefont {T.}~\bibnamefont
  {Liang}}, \bibinfo {author} {\bibfnamefont {J.}~\bibnamefont {Lin}}, \bibinfo
  {author} {\bibfnamefont {Q.}~\bibnamefont {Gibson}}, \bibinfo {author}
  {\bibfnamefont {S.}~\bibnamefont {Kushwaha}}, \bibinfo {author}
  {\bibfnamefont {M.}~\bibnamefont {Liu}}, \bibinfo {author} {\bibfnamefont
  {W.}~\bibnamefont {Wang}}, \bibinfo {author} {\bibfnamefont {H.}~\bibnamefont
  {Xiong}}, \bibinfo {author} {\bibfnamefont {J.~A.}\ \bibnamefont {Sobota}},
  \bibinfo {author} {\bibfnamefont {M.}~\bibnamefont {Hashimoto}}, \bibinfo
  {author} {\bibfnamefont {P.~S.}\ \bibnamefont {Kirchmann}}, \emph {et~al.},\
  }\bibfield  {title} {\bibinfo {title} {Anomalous {Hall} effect in {ZrTe5}},\
  }\href@noop {} {\bibfield  {journal} {\bibinfo  {journal} {Nat. Phys.}\
  }\textbf {\bibinfo {volume} {14}},\ \bibinfo {pages} {451} (\bibinfo {year}
  {2018})}\BibitemShut {NoStop}%
\bibitem [{\citenamefont {McIver}\ \emph {et~al.}(2020)\citenamefont {McIver},
  \citenamefont {Schulte}, \citenamefont {Stein}, \citenamefont {Matsuyama},
  \citenamefont {Jotzu}, \citenamefont {Meier},\ and\ \citenamefont
  {Cavalleri}}]{mciver2020light}%
  \BibitemOpen
  \bibfield  {author} {\bibinfo {author} {\bibfnamefont {J.~W.}\ \bibnamefont
  {McIver}}, \bibinfo {author} {\bibfnamefont {B.}~\bibnamefont {Schulte}},
  \bibinfo {author} {\bibfnamefont {F.-U.}\ \bibnamefont {Stein}}, \bibinfo
  {author} {\bibfnamefont {T.}~\bibnamefont {Matsuyama}}, \bibinfo {author}
  {\bibfnamefont {G.}~\bibnamefont {Jotzu}}, \bibinfo {author} {\bibfnamefont
  {G.}~\bibnamefont {Meier}},\ and\ \bibinfo {author} {\bibfnamefont
  {A.}~\bibnamefont {Cavalleri}},\ }\bibfield  {title} {\bibinfo {title}
  {Light-induced anomalous {Hall} effect in graphene},\ }\href@noop {}
  {\bibfield  {journal} {\bibinfo  {journal} {Nat. Phys.}\ }\textbf {\bibinfo
  {volume} {16}},\ \bibinfo {pages} {38} (\bibinfo {year} {2020})}\BibitemShut
  {NoStop}%
\bibitem [{\citenamefont {Laughlin}(1983)}]{laughlin1983anomalous}%
  \BibitemOpen
  \bibfield  {author} {\bibinfo {author} {\bibfnamefont {R.~B.}\ \bibnamefont
  {Laughlin}},\ }\bibfield  {title} {\bibinfo {title} {Anomalous quantum {Hall}
  effect: an incompressible quantum fluid with fractionally charged
  excitations},\ }\href@noop {} {\bibfield  {journal} {\bibinfo  {journal}
  {Phys. Rev. Lett.}\ }\textbf {\bibinfo {volume} {50}},\ \bibinfo {pages}
  {1395} (\bibinfo {year} {1983})}\BibitemShut {NoStop}%
\bibitem [{\citenamefont {Yu}\ \emph {et~al.}(2010)\citenamefont {Yu},
  \citenamefont {Zhang}, \citenamefont {Zhang}, \citenamefont {Zhang},
  \citenamefont {Dai},\ and\ \citenamefont {Fang}}]{yu2010quantized}%
  \BibitemOpen
  \bibfield  {author} {\bibinfo {author} {\bibfnamefont {R.}~\bibnamefont
  {Yu}}, \bibinfo {author} {\bibfnamefont {W.}~\bibnamefont {Zhang}}, \bibinfo
  {author} {\bibfnamefont {H.-J.}\ \bibnamefont {Zhang}}, \bibinfo {author}
  {\bibfnamefont {S.-C.}\ \bibnamefont {Zhang}}, \bibinfo {author}
  {\bibfnamefont {X.}~\bibnamefont {Dai}},\ and\ \bibinfo {author}
  {\bibfnamefont {Z.}~\bibnamefont {Fang}},\ }\bibfield  {title} {\bibinfo
  {title} {Quantized anomalous {Hall} effect in magnetic topological
  insulators},\ }\href@noop {} {\bibfield  {journal} {\bibinfo  {journal}
  {Science}\ }\textbf {\bibinfo {volume} {329}},\ \bibinfo {pages} {61}
  (\bibinfo {year} {2010})}\BibitemShut {NoStop}%
\bibitem [{\citenamefont {Chang}\ \emph {et~al.}(2013)\citenamefont {Chang},
  \citenamefont {Zhang}, \citenamefont {Feng}, \citenamefont {Shen},
  \citenamefont {Zhang}, \citenamefont {Guo}, \citenamefont {Li}, \citenamefont
  {Ou}, \citenamefont {Wei}, \citenamefont {Wang} \emph
  {et~al.}}]{chang2013experimental}%
  \BibitemOpen
  \bibfield  {author} {\bibinfo {author} {\bibfnamefont {C.-Z.}\ \bibnamefont
  {Chang}}, \bibinfo {author} {\bibfnamefont {J.}~\bibnamefont {Zhang}},
  \bibinfo {author} {\bibfnamefont {X.}~\bibnamefont {Feng}}, \bibinfo {author}
  {\bibfnamefont {J.}~\bibnamefont {Shen}}, \bibinfo {author} {\bibfnamefont
  {Z.}~\bibnamefont {Zhang}}, \bibinfo {author} {\bibfnamefont
  {M.}~\bibnamefont {Guo}}, \bibinfo {author} {\bibfnamefont {K.}~\bibnamefont
  {Li}}, \bibinfo {author} {\bibfnamefont {Y.}~\bibnamefont {Ou}}, \bibinfo
  {author} {\bibfnamefont {P.}~\bibnamefont {Wei}}, \bibinfo {author}
  {\bibfnamefont {L.-L.}\ \bibnamefont {Wang}}, \emph {et~al.},\ }\bibfield
  {title} {\bibinfo {title} {Experimental observation of the quantum anomalous
  {Hall} effect in a magnetic topological insulator},\ }\href@noop {}
  {\bibfield  {journal} {\bibinfo  {journal} {Science}\ }\textbf {\bibinfo
  {volume} {340}},\ \bibinfo {pages} {167} (\bibinfo {year}
  {2013})}\BibitemShut {NoStop}%
\bibitem [{\citenamefont {Deng}\ \emph {et~al.}(2020)\citenamefont {Deng},
  \citenamefont {Yu}, \citenamefont {Shi}, \citenamefont {Guo}, \citenamefont
  {Xu}, \citenamefont {Wang}, \citenamefont {Chen},\ and\ \citenamefont
  {Zhang}}]{deng2020quantum}%
  \BibitemOpen
  \bibfield  {author} {\bibinfo {author} {\bibfnamefont {Y.}~\bibnamefont
  {Deng}}, \bibinfo {author} {\bibfnamefont {Y.}~\bibnamefont {Yu}}, \bibinfo
  {author} {\bibfnamefont {M.~Z.}\ \bibnamefont {Shi}}, \bibinfo {author}
  {\bibfnamefont {Z.}~\bibnamefont {Guo}}, \bibinfo {author} {\bibfnamefont
  {Z.}~\bibnamefont {Xu}}, \bibinfo {author} {\bibfnamefont {J.}~\bibnamefont
  {Wang}}, \bibinfo {author} {\bibfnamefont {X.~H.}\ \bibnamefont {Chen}},\
  and\ \bibinfo {author} {\bibfnamefont {Y.}~\bibnamefont {Zhang}},\ }\bibfield
   {title} {\bibinfo {title} {Quantum anomalous {Hall} effect in intrinsic
  magnetic topological insulator {MnBi2Te4}},\ }\href@noop {} {\bibfield
  {journal} {\bibinfo  {journal} {Science}\ }\textbf {\bibinfo {volume}
  {367}},\ \bibinfo {pages} {895} (\bibinfo {year} {2020})}\BibitemShut
  {NoStop}%
\bibitem [{\citenamefont {Chang}\ \emph {et~al.}(2023)\citenamefont {Chang},
  \citenamefont {Liu},\ and\ \citenamefont {MacDonald}}]{chang2023colloquium}%
  \BibitemOpen
  \bibfield  {author} {\bibinfo {author} {\bibfnamefont {C.-Z.}\ \bibnamefont
  {Chang}}, \bibinfo {author} {\bibfnamefont {C.-X.}\ \bibnamefont {Liu}},\
  and\ \bibinfo {author} {\bibfnamefont {A.~H.}\ \bibnamefont {MacDonald}},\
  }\bibfield  {title} {\bibinfo {title} {Colloquium: Quantum anomalous {Hall}
  effect},\ }\href@noop {} {\bibfield  {journal} {\bibinfo  {journal} {Rev.
  Mod. Phys.}\ }\textbf {\bibinfo {volume} {95}},\ \bibinfo {pages} {011002}
  (\bibinfo {year} {2023})}\BibitemShut {NoStop}%
\bibitem [{\citenamefont {Born}\ and\ \citenamefont
  {Huang}(1996)}]{born1996dynamical}%
  \BibitemOpen
  \bibfield  {author} {\bibinfo {author} {\bibfnamefont {M.}~\bibnamefont
  {Born}}\ and\ \bibinfo {author} {\bibfnamefont {K.}~\bibnamefont {Huang}},\
  }\href@noop {} {\emph {\bibinfo {title} {Dynamical theory of crystal
  lattices}}}\ (\bibinfo  {publisher} {Oxford university press},\ \bibinfo
  {year} {1996})\BibitemShut {NoStop}%
\bibitem [{\citenamefont {Mead}\ and\ \citenamefont
  {Truhlar}(1979)}]{mead1979determination}%
  \BibitemOpen
  \bibfield  {author} {\bibinfo {author} {\bibfnamefont {C.~A.}\ \bibnamefont
  {Mead}}\ and\ \bibinfo {author} {\bibfnamefont {D.~G.}\ \bibnamefont
  {Truhlar}},\ }\bibfield  {title} {\bibinfo {title} {On the determination of
  {Born--Oppenheimer} nuclear motion wave functions including complications due
  to conical intersections and identical nuclei},\ }\href@noop {} {\bibfield
  {journal} {\bibinfo  {journal} {J. Chem. Phys.}\ }\textbf {\bibinfo {volume}
  {70}},\ \bibinfo {pages} {2284} (\bibinfo {year} {1979})}\BibitemShut
  {NoStop}%
\bibitem [{\citenamefont {Mead}(1992)}]{mead1992geometric}%
  \BibitemOpen
  \bibfield  {author} {\bibinfo {author} {\bibfnamefont {C.~A.}\ \bibnamefont
  {Mead}},\ }\bibfield  {title} {\bibinfo {title} {The geometric phase in
  molecular systems},\ }\href@noop {} {\bibfield  {journal} {\bibinfo
  {journal} {Rev. Mod. Phys.}\ }\textbf {\bibinfo {volume} {64}},\ \bibinfo
  {pages} {51} (\bibinfo {year} {1992})}\BibitemShut {NoStop}%
\bibitem [{\citenamefont {Min}\ \emph {et~al.}(2014)\citenamefont {Min},
  \citenamefont {Abedi}, \citenamefont {Kim},\ and\ \citenamefont
  {Gross}}]{min2014molecular}%
  \BibitemOpen
  \bibfield  {author} {\bibinfo {author} {\bibfnamefont {S.~K.}\ \bibnamefont
  {Min}}, \bibinfo {author} {\bibfnamefont {A.}~\bibnamefont {Abedi}}, \bibinfo
  {author} {\bibfnamefont {K.~S.}\ \bibnamefont {Kim}},\ and\ \bibinfo {author}
  {\bibfnamefont {E.}~\bibnamefont {Gross}},\ }\bibfield  {title} {\bibinfo
  {title} {Is the molecular {Berry} phase an artifact of the {Born-Oppenheimer}
  approximation?},\ }\href@noop {} {\bibfield  {journal} {\bibinfo  {journal}
  {Phys. Rev. Lett.}\ }\textbf {\bibinfo {volume} {113}},\ \bibinfo {pages}
  {263004} (\bibinfo {year} {2014})}\BibitemShut {NoStop}%
\bibitem [{\citenamefont {Saparov}\ \emph {et~al.}(2022)\citenamefont
  {Saparov}, \citenamefont {Xiong}, \citenamefont {Ren},\ and\ \citenamefont
  {Niu}}]{saparov2022lattice}%
  \BibitemOpen
  \bibfield  {author} {\bibinfo {author} {\bibfnamefont {D.}~\bibnamefont
  {Saparov}}, \bibinfo {author} {\bibfnamefont {B.}~\bibnamefont {Xiong}},
  \bibinfo {author} {\bibfnamefont {Y.}~\bibnamefont {Ren}},\ and\ \bibinfo
  {author} {\bibfnamefont {Q.}~\bibnamefont {Niu}},\ }\bibfield  {title}
  {\bibinfo {title} {Lattice dynamics with molecular {Berry} curvature:
  {Chiral} optical phonons},\ }\href@noop {} {\bibfield  {journal} {\bibinfo
  {journal} {Phys. Rev. B}\ }\textbf {\bibinfo {volume} {105}},\ \bibinfo
  {pages} {064303} (\bibinfo {year} {2022})}\BibitemShut {NoStop}%
\bibitem [{\citenamefont {Bonini}\ \emph {et~al.}(2023)\citenamefont {Bonini},
  \citenamefont {Ren}, \citenamefont {Vanderbilt}, \citenamefont {Stengel},
  \citenamefont {Dreyer},\ and\ \citenamefont {Coh}}]{PhysRevLett.130.086701}%
  \BibitemOpen
  \bibfield  {author} {\bibinfo {author} {\bibfnamefont {J.}~\bibnamefont
  {Bonini}}, \bibinfo {author} {\bibfnamefont {S.}~\bibnamefont {Ren}},
  \bibinfo {author} {\bibfnamefont {D.}~\bibnamefont {Vanderbilt}}, \bibinfo
  {author} {\bibfnamefont {M.}~\bibnamefont {Stengel}}, \bibinfo {author}
  {\bibfnamefont {C.~E.}\ \bibnamefont {Dreyer}},\ and\ \bibinfo {author}
  {\bibfnamefont {S.}~\bibnamefont {Coh}},\ }\bibfield  {title} {\bibinfo
  {title} {Frequency splitting of chiral phonons from broken time-reversal
  symmetry in {CrI3}},\ }\href@noop {} {\bibfield  {journal} {\bibinfo
  {journal} {Phys. Rev. Lett.}\ }\textbf {\bibinfo {volume} {130}},\ \bibinfo
  {pages} {086701} (\bibinfo {year} {2023})}\BibitemShut {NoStop}%
\bibitem [{sup()}]{supplemental}%
  \BibitemOpen
  \href@noop {} {}\bibinfo {howpublished} {See Supplementary Material at [] for
  the calculation method, symmetry constraints, other component of molecular
  berry curvature and phonon properties without molecular berry
  curvature.}\BibitemShut {Stop}%
\bibitem [{\citenamefont {Zhang}\ and\ \citenamefont
  {Niu}(2014)}]{zhang2014angular}%
  \BibitemOpen
  \bibfield  {author} {\bibinfo {author} {\bibfnamefont {L.}~\bibnamefont
  {Zhang}}\ and\ \bibinfo {author} {\bibfnamefont {Q.}~\bibnamefont {Niu}},\
  }\bibfield  {title} {\bibinfo {title} {Angular momentum of phonons and the
  {Einstein--de Haas} effect},\ }\href@noop {} {\bibfield  {journal} {\bibinfo
  {journal} {Phys. Rev. Lett.}\ }\textbf {\bibinfo {volume} {112}},\ \bibinfo
  {pages} {085503} (\bibinfo {year} {2014})}\BibitemShut {NoStop}%
\bibitem [{\citenamefont {Zhang}\ and\ \citenamefont
  {Niu}(2015)}]{zhang2015chiral}%
  \BibitemOpen
  \bibfield  {author} {\bibinfo {author} {\bibfnamefont {L.}~\bibnamefont
  {Zhang}}\ and\ \bibinfo {author} {\bibfnamefont {Q.}~\bibnamefont {Niu}},\
  }\bibfield  {title} {\bibinfo {title} {Chiral phonons at high-symmetry points
  in monolayer hexagonal lattices},\ }\href@noop {} {\bibfield  {journal}
  {\bibinfo  {journal} {Phys. Rev. Lett.}\ }\textbf {\bibinfo {volume} {115}},\
  \bibinfo {pages} {115502} (\bibinfo {year} {2015})}\BibitemShut {NoStop}%
\bibitem [{\citenamefont {Wang}\ \emph {et~al.}(2022)\citenamefont {Wang},
  \citenamefont {Li}, \citenamefont {Zhu}, \citenamefont {Chen}, \citenamefont
  {Wu}, \citenamefont {Gao}, \citenamefont {Zhang},\ and\ \citenamefont
  {Yang}}]{wang2022chiral}%
  \BibitemOpen
  \bibfield  {author} {\bibinfo {author} {\bibfnamefont {Q.}~\bibnamefont
  {Wang}}, \bibinfo {author} {\bibfnamefont {S.}~\bibnamefont {Li}}, \bibinfo
  {author} {\bibfnamefont {J.}~\bibnamefont {Zhu}}, \bibinfo {author}
  {\bibfnamefont {H.}~\bibnamefont {Chen}}, \bibinfo {author} {\bibfnamefont
  {W.}~\bibnamefont {Wu}}, \bibinfo {author} {\bibfnamefont {W.}~\bibnamefont
  {Gao}}, \bibinfo {author} {\bibfnamefont {L.}~\bibnamefont {Zhang}},\ and\
  \bibinfo {author} {\bibfnamefont {S.~A.}\ \bibnamefont {Yang}},\ }\bibfield
  {title} {\bibinfo {title} {Chiral phonons in lattices with {C4} symmetry},\
  }\href@noop {} {\bibfield  {journal} {\bibinfo  {journal} {Phys. Rev. B}\
  }\textbf {\bibinfo {volume} {105}},\ \bibinfo {pages} {104301} (\bibinfo
  {year} {2022})}\BibitemShut {NoStop}%
\bibitem [{\citenamefont {Zhu}\ \emph {et~al.}(2018)\citenamefont {Zhu},
  \citenamefont {Yi}, \citenamefont {Li}, \citenamefont {Xiao}, \citenamefont
  {Zhang}, \citenamefont {Yang}, \citenamefont {Kaindl}, \citenamefont {Li},
  \citenamefont {Wang},\ and\ \citenamefont {Zhang}}]{zhu2018observation}%
  \BibitemOpen
  \bibfield  {author} {\bibinfo {author} {\bibfnamefont {H.}~\bibnamefont
  {Zhu}}, \bibinfo {author} {\bibfnamefont {J.}~\bibnamefont {Yi}}, \bibinfo
  {author} {\bibfnamefont {M.-Y.}\ \bibnamefont {Li}}, \bibinfo {author}
  {\bibfnamefont {J.}~\bibnamefont {Xiao}}, \bibinfo {author} {\bibfnamefont
  {L.}~\bibnamefont {Zhang}}, \bibinfo {author} {\bibfnamefont {C.-W.}\
  \bibnamefont {Yang}}, \bibinfo {author} {\bibfnamefont {R.~A.}\ \bibnamefont
  {Kaindl}}, \bibinfo {author} {\bibfnamefont {L.-J.}\ \bibnamefont {Li}},
  \bibinfo {author} {\bibfnamefont {Y.}~\bibnamefont {Wang}},\ and\ \bibinfo
  {author} {\bibfnamefont {X.}~\bibnamefont {Zhang}},\ }\bibfield  {title}
  {\bibinfo {title} {Observation of chiral phonons},\ }\href@noop {} {\bibfield
   {journal} {\bibinfo  {journal} {Science}\ }\textbf {\bibinfo {volume}
  {359}},\ \bibinfo {pages} {579} (\bibinfo {year} {2018})}\BibitemShut
  {NoStop}%
\bibitem [{\citenamefont {Flebus}\ and\ \citenamefont
  {MacDonald}(2023)}]{PhysRevLett.131.236301}%
  \BibitemOpen
  \bibfield  {author} {\bibinfo {author} {\bibfnamefont {B.}~\bibnamefont
  {Flebus}}\ and\ \bibinfo {author} {\bibfnamefont {A.~H.}\ \bibnamefont
  {MacDonald}},\ }\bibfield  {title} {\bibinfo {title} {Phonon {Hall} viscosity
  of ionic crystals},\ }\href@noop {} {\bibfield  {journal} {\bibinfo
  {journal} {Phys. Rev. Lett.}\ }\textbf {\bibinfo {volume} {131}},\ \bibinfo
  {pages} {236301} (\bibinfo {year} {2023})}\BibitemShut {NoStop}%
\end{thebibliography}%

\end{document}